\documentclass[pra,aps,superscriptaddress,twocolumn,notitlepage,showpacs]{revtex4-2}
\usepackage{latexsym}
\usepackage{amsmath}
\usepackage{amssymb}
\usepackage{graphicx}
\usepackage{caption}
\usepackage{subfigure}
\usepackage{float}
\usepackage{mathrsfs}
\usepackage{xcolor}
\usepackage{color}
\usepackage{txfonts}
\usepackage[justification=centering,
            format=plain]{caption}
\usepackage{multirow}
\usepackage{booktabs}
\usepackage{amsmath}
\usepackage{verbatim}

\renewcommand{\raggedright}{\leftskip=0pt \rightskip=0pt plus 0cm}
\begin{document}

\title{Non-invasive imaging of object behind strongly scattering  media via cross-spectrum}

\author{Xingchen Zhao}
\affiliation{Texas A\&M University, College Station, TX 77843, USA}

\author{Tao Peng}
\email{taopeng@tamu.edu}
\affiliation{Texas A\&M University, College Station, TX 77843, USA}

\author{Zhenhuan Yi}
\affiliation{Texas A\&M University, College Station, TX 77843, USA}

\author{Lida Zhang}
\affiliation{Texas A\&M University, College Station, TX 77843, USA}
\affiliation{Aarhus University, 8000 Aarhus C, Denmark}

\author{M. Suhail Zubairy}
\affiliation{Texas A\&M University, College Station, TX 77843, USA}

\author{Yanhua Shih}
\affiliation{University of Maryland, Baltimore County, Baltimore, Maryland 21250, USA} 

\author{Marlan O. Scully}
\affiliation{Texas A\&M University, College Station, TX 77843, USA}
\affiliation{Baylor University, Waco, TX 76706, USA}
\affiliation{Princeton University, Princeton, NJ 08544, USA}

\date{\today}

\begin{abstract}
We develop a method based on the cross-spectrum of an intensity-modulated CW laser, which can extract a signal from an extremely noisy environment and image objects hidden in strongly scattering media. We theoretically analyzed our scheme and performed the experiment by scanning the object placed in between two ground glass diffusers. The image of the object is retrieved by collecting the amplitudes at the modulation frequency of all the cross-spectra. Our method is non-invasive, easy-to-implement, and can work for both static and dynamic media. 
\end{abstract}

\maketitle

\section{Introduction}
Optical observation through scattering media is a difficult task in optics~\cite{meyers2011turbulence, mosk2012controlling, bertolotti2012non, rotter2017light, ntziachristos2010going,bhattacharjee2020controlling}. Imaging through strongly scattering  (visually opaque) media is especially challenging. The randomization of optical wavefront due to strong scattering scrambles the spatial information and smears the images obtained by light intensity measurement. A variety of strategies have been devised to image objects hidden behind the strongly scattering media. Some methods attempt to extract the non-scattered photons, such as time-gating~\cite{wang1991ballistic, hee1993femtosecond, das1993ultrafast, liu1994transmitted}, coherence-gating~\cite{abramson1989single, chen1991two,leith1991electronic}, and rotating polarization methods~\cite{ramachandran1998two, emile1996rotating}. These techniques suffer from low signal-to-noise ratio due to the tiny amount of non-scattered photons, which greatly limits the imaging (or penetration) depth. In addition, some gating techniques require the use of ultrafast laser pulses, which may be devastating for living biological tissue. Other methods focus on reversing the scattering process and recover input optical information directly from scattered photons, such as optical phase conjugation~\cite{cui2010implementation, yaqoob2008optical}, transmission matrix~\cite{popoff2010measuring,yoon2015measuring, mounaix2016deterministic,deaguiarEnhancedNonlinearImaging2016}, and speckle correlation~\cite{katz2014non,newman2016imaging,newman2016imaging,  webb2020theory,luoParametrizationSpeckleIntensity2020}. However, these methods are either invasive or require intensive computations using iterative algorithms that can only work for static media. Imaging through dynamic media is still quite challenging~\cite{yuan2017non, sun2019image, ruan2020fluorescence}. In this case, the time-dependent mapping between input and output fields requires instant completion of the image reconstruction process to follow the variation of media. Therefore, the capability of iterative algorithms is greatly reduced.

In this letter, we report a method based on the cross-spectrum measurement from two single-pixel detectors with an intensity-modulated CW laser. The cross-spectrum technique has been mainly used to analyze the cross-correlation between two time series in the frequency domain. Intensity-modulated CW laser was widely used in diffuse optical imaging to probe optical properties in living tissue~\cite{fishkin1991diffusion, tromberg1991optical, o2012diffuse}. We adopt these techniques to demonstrate a non-invasive and easy-to-implement scheme, by which the image of an object can be reconstructed not only through both static and dynamic diffusers but also under extremely noisy environment, \textit{i.e.}, the light intensity is much lower than detector noise. Besides, the use of CW laser makes the method more favorable in applications involving living tissues. 

\section{Theoretical description of the method}
\begin{figure}[ht]
	\centering
	\includegraphics[width=\linewidth]{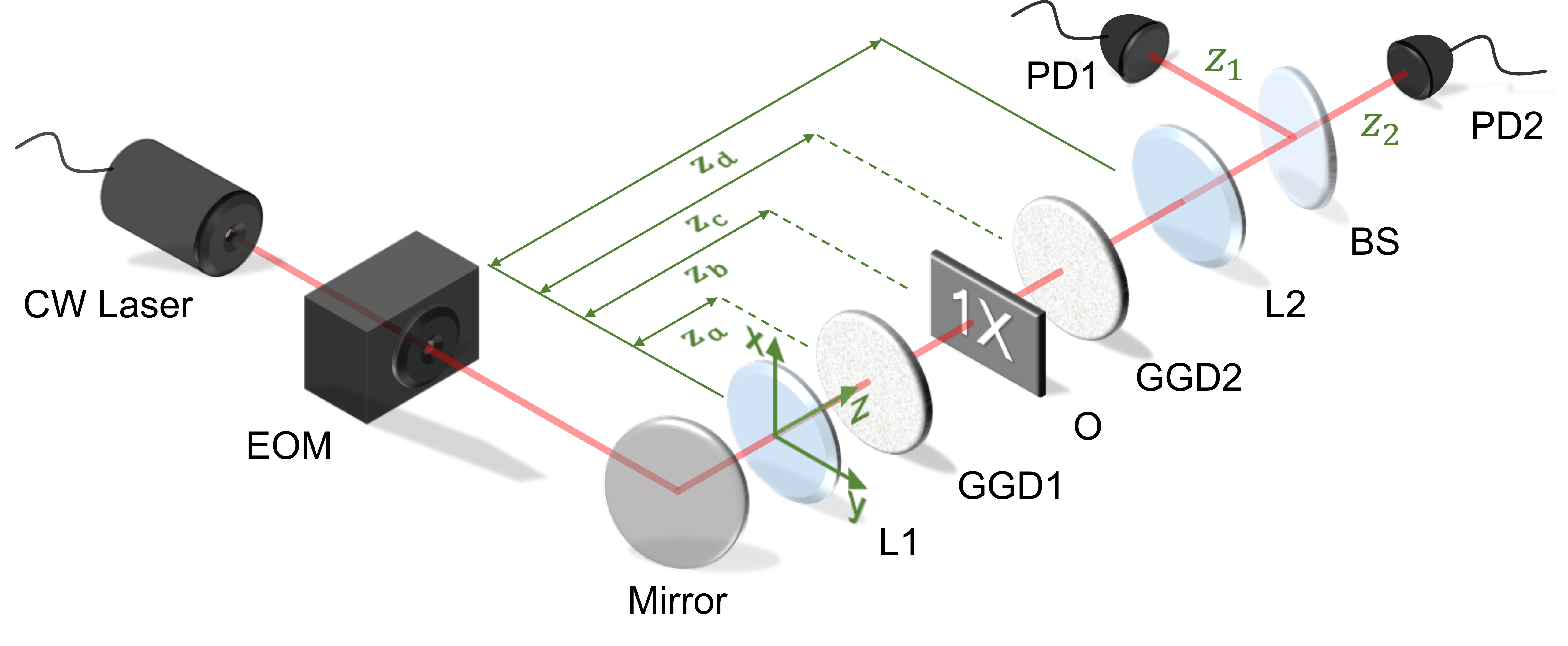}
	\captionsetup{justification=raggedright, singlelinecheck=false}
	\caption{Schematic of the experimental setup. We use a modulated CW laser for illumination. The object is sandwiched between two GGDs. Signal at each detector is made to be much lower than the noise level.  EOM: electro-optic modulator; L: lens; GGD: ground glass diffuser; O: object; BS: beam splitter; PD: photodetector. The Cartesian coordinate is located in the center of L1 with $z$-axis pointing along the propagation direction of the light. $z_{1}$ and $z_{2}$ are the distances between L1 and the two PDs, respectively.}
	\label{fig:fig_1_exp_setup}
\end{figure}

The experimental setup is shown in Fig.~\ref{fig:fig_1_exp_setup}. A CW laser (633 nm, QPhotonics, QFBGLD-633-30PM) is intensity-modulated by an electro-optic modulator (EOM: Thorlabs, EO-AM-NR-C1) at frequency $f_{mod}=1$ MHz. An objective lens (L1: Nikon, plan fluor, 10X/0.30, $\infty$/0, WD 17.5)
is used to focus the modulated light onto the object plate (O: Thorlabs, R3L1S4N resolution test target) where the letter ``1X'' is transparent (height: $\sim 2.3$ mm; width: $\sim 3.3$ mm; width of transparent region: $\sim 0.36$ mm). The object plate is sandwiched in situ between a pair of ground glass diffusers (GGDs: Thorlabs, DG10-220) of 220 grit (average grit diameter $\bar{d}_{grit}=53\ \mu\text{m}$). The GGD serves as the strongly scattering medium in our experiment, as widely used in variety of imaging scenarios~\cite{suzuki2014frequency, roy2016analysis, wu2016single, tzang2019wavefront}. The focal spot is $\sim 2.7 ~\mu$m in size without the GGD, and is estimated to be $\sim 25 ~\mu$m when GGD1 is present. A second lens (L2: $f=150$ mm) is placed behind GDD2 to collect the scattered light. The distance between the object and each diffuser is $\sim 5$ mm (we note here that the distance cannot be too small due to the shower-curtain effect \cite{edrei2016optical}). The GGDs can be either kept static or moved back and forth together by a motorized stage. The output light is split into two arms by a beam splitter (BS), which are then measured by two photodetectors (PDs: Thorlabs, PDA 10A) respectively, where the two PDs are put at the focal plane of the lens. The data is then sent to a computer to generate images of the object. The object is scanned pixel-by-pixel with an appropriate step size to resolve the region of interest. We note here that, due to the low incident laser power ($\sim 2~\mu W$) and strong scattering from the two GGDs ($\sim 75~nW$ at the detector plane), the laser power measured at each PD is buried in the electronic and environmental noise.  

We first outline a brief theoretical description of the cross-spectrum method~\cite{estes1971scattering, churnside1982speckle, goodman2007speckle, foley1978directionality}. As shown in Fig.~\ref{fig:fig_1_exp_setup}, a Cartesian coordinate system is placed in the center of L1, with the $z$-axis pointing along the propagation direction of the light. A collimated incident beam of radius $\sigma$ is focused by L1 with focal length $f_{1}$. The scattering centers on GGD1 will produce an electric field at distance $z$ with the form
\begin{align}\label{eq: field_scattered}
& E\left(\boldsymbol{\rho}_{z},z,t\right)=\frac{-ik}{2\pi}A\left(z-z_{a}\right)\times\nonumber \\
 & \qquad\int\text{d}^{2}\boldsymbol{\rho}_{z_{a}}E\left(\boldsymbol{\rho}_{z_{a}},z_{a},t\right)R\left(\boldsymbol{\rho}_{z_{a}}\right)G\left(\boldsymbol{\rho}_{z}-\boldsymbol{\rho}_{z_{a}}; z-z_{a}\right),
\end{align}
where $\boldsymbol{\rho}_{z}=\left(x_{z},y_{z}\right)$ is the position vector in the plane at distance $z$ from L1, $k$ is the wave vector, $R\left(\boldsymbol{\rho}_{z_{a}}\right)$ describes GGD1 as a phase plate due to the scattering centers at $\boldsymbol{\rho}_{z_a}$, which imprint the random phase profile on the propagating field. $E\left(\boldsymbol{\rho}_{z_{a}},z_{a},t\right)$ is the profile of the field at the left surface of GGD1, which is given by
\begin{align}\label{eq: field_profile}
 E\left(\boldsymbol{\rho}_{z_{a}},z_{a},t\right) & =E_{a}E_{0}\left(t\right)e^{-i(\nu_{0}t-kz_{a})}\cr
 & \times\exp\left\{ -\frac{E_{a}}{2}\left(\frac{ik}{f_{1}}+\frac{1}{2\sigma^{2}}\right)\boldsymbol{\rho}_{z_{a}}^{2}\right\},
\end{align}
where $E_a=-\frac{ik}{2z_a}\frac{1}{1/4\sigma^{2}+\left(ik/2\right)\left(1/f-1/z_a\right)}$, $E_{0}\left(t\right)=\sqrt{I_{0}\cos2\pi f_{mod}t}$ expresses a sinusoidal-modulation of light intensity $I_{0}$ at frequency $f_{mod}$, and $\nu_{0}$ is the frequency of the laser. We also define $A\left(z\right)=e^{ikz}/z$ and  $G\left(\boldsymbol{\alpha};\beta\right)=e^{ik\left|\boldsymbol{\alpha}\right|^{2}/2\beta}$. $\boldsymbol{\rho}_{z}=\left(x_{z},y_{z}\right)$ is the position vector in the receiver plane at distance $z$ from L1. The integration in Eq.~(\ref{eq: field_scattered}) is performed over the illumination area on GGD1. 

When the step size is made roughly about the same size as the focal spot after GGD1, and much smaller as compared to the object size, the object is considered as scanned point by point, \textit{i.e.}, the sample transparency can be considered constant for each scanning point. We can model the object as a transmission function $T\left(\boldsymbol{\rho}_{z_{b}}\right)$ where $\boldsymbol{\rho}_{z_{b}}=\left(x_{z_{b}},y_{z_{b}}\right)$ is the position vector in the object plane. Upon passing through the object, being scattered by GGD2, and being collected by L2, the field at the two detectors is found to be
\begin{align}\label{eq: field_at_PDs}
E\left(\boldsymbol{\rho}_{z_{b}},z_{j},t\right) & =\frac{-ik}{2\pi}\tilde{A}\iiiint\text{d}^{2}\boldsymbol{\rho}_{z_{j}}\text{d}^{2}\boldsymbol{\rho}_{z_{d}}\text{d}^{2}\boldsymbol{\rho}_{z_{c}}\text{d}^{2}\boldsymbol{\rho}_{z_{a}}\nonumber \\
& \times E\left(\boldsymbol{\rho}_{z_{a}},z_{a},t\right)R\left(\boldsymbol{\rho}_{z_{a}}\right)R\left(\boldsymbol{\rho}_{z_{c}}\right)T\left(\boldsymbol{\rho}_{z_{b}}\right)\nonumber \\
& \times G\left(\boldsymbol{\rho}_{z_{c}}-\boldsymbol{\rho}_{z_{a}},z_{c}-z_{a}\right)G\left(\boldsymbol{\rho}_{z_{d}}-\boldsymbol{\rho}_{z_{c}},z_{d}-z_{c}\right)\nonumber \\
& \times G\left(\boldsymbol{\rho}_{z_{j}}-\boldsymbol{\rho}_{z_{d}},z_{j}-z_{d}\right)G\left(-\boldsymbol{\rho}_{z_{d}},f_{2}\right),
\end{align}
where $\tilde{A}=A\left(z_{c}-z_{a}\right)A\left(z_{d}-z_{c}\right)A\left(z_{j}-z_{d}\right)
$, $z_{j}$ ($j=1,2$) denote the distance between the detector $j$ and L1, $f_{2}$ is the focal length of L2, $R\left(\boldsymbol{\rho}_{z_{c}}\right)$ describes the random phase profile due to scattering on GGD2. $G\left(-\boldsymbol{\rho}_{z_{d}}; f_{2}\right)$ is the propagation factor of L2. 

The total signals measured by the two photodetectors for each scanning position centered at $\boldsymbol{\rho}_{z_{b}}$ can be expressed as
\begin{equation}\label{eq: signal_at_detectors}
    S\left(\boldsymbol{\rho}_{z_{b}},z_{j},t\right)=I\left(\boldsymbol{\rho}_{z_{b}},z_{j},t\right)+\beta_{j}N\left(t\right),
\end{equation}
where $I\left(\boldsymbol{\rho}_{z_{b}},z_{j},t\right)\equiv E\left(\boldsymbol{\rho}_{z_{b}},z_{j},t\right)E^{\ast}\left(\boldsymbol{\rho}_{z_{b}},z_{j},t\right)$ is the intensity at detector $j$, $N\left(t\right)$ is a white noise distribution that models all the noise due to detectors and environment, and $\beta_{j}$ is the amplitude of the noise at detector $j$. It follows that the time-domain cross-correlation is given by
\begin{equation}\label{eq: c_tau_def}
    C\left(\boldsymbol{\rho}_{z_{b}},\tau\right)=\left\langle \int_{0}^{\mathcal{T}}\text{d}tS^{\ast}\left(\boldsymbol{\rho}_{z_{b}},z_{1},t\right)S\left(\boldsymbol{\rho}_{z_{b}},z_{2},t+\tau\right)\right\rangle,
\end{equation}
where $\mathcal{T}$ is the measurement time. 
We further assumed that the correlations between intensity and noise vanish since they are uncorrelated. On substituting from Eq.~(\ref{eq: signal_at_detectors}) into Eq.~(\ref{eq: c_tau_def}), we obtain
\begin{align}\label{eq: c_tau_nonvanishing}
C\left(\boldsymbol{\rho}_{z_{b}},\tau\right) & =\int_{0}^{\mathcal{T}}\text{d}t\left\langle I\left(\boldsymbol{\rho}_{z_{b}},z_{1},t\right)I\left(\boldsymbol{\rho}_{z_{b}},z_{2},t+\tau\right)\right\rangle \nonumber \\
& +\beta_{1}\beta_{2}\int_{0}^{\mathcal{T}}\text{d}t\left\langle N\left(t\right)N\left(t+\tau\right)\right\rangle.
\end{align} 
The intensity correlation in the first term of Eq.~(\ref{eq: c_tau_nonvanishing}) can be expressed as
\begin{align}\label{eq: cc_I}
 & \left\langle I\left(\boldsymbol{\rho}_{z_{b}},z_{1},t\right)I\left(\boldsymbol{\rho}_{z_{b}},z_{2},t+\tau\right)\right\rangle \nonumber \\
 & =\left\langle E\left(\boldsymbol{\rho}_{z_{b}},z_{1},t\right)E^{\ast}\left(\boldsymbol{\rho}_{z_{b}},z_{1},t\right)E\left(\boldsymbol{\rho}_{z_{b}},z_{2},t+\tau\right)E^{\ast}\left(\boldsymbol{\rho}_{z_{b}},z_{2},t+\tau\right)\right\rangle 
\end{align}
where $E\left(\boldsymbol{\rho}_{z_{b}},z_{j},t\right)$ $(j=1,2)$ is given by Eq.~(\ref{eq: field_at_PDs}).
Since the scattering centers are independent of each other and satisfy Gaussian statistics, the random phase term $R\left(\boldsymbol{\rho}_{z_{i}}\right)$ obeys \begin{equation}\label{eq: corr_random_phases}
    \left\langle R\left(\boldsymbol{\rho}_{z_{i}}\right)R^{\ast}\left(\boldsymbol{\rho}_{z_{i}}^{\prime}\right)\right\rangle =\delta\left(\boldsymbol{\rho}_{z_{i}}-\boldsymbol{\rho}_{z_{i}}^{\prime}\right)
\end{equation} and
\begin{align}\label{eq: avg_random_phases}
 & \quad\left\langle R\left(\boldsymbol{\rho}_{z_{i}}\right)R^{\ast}\left(\boldsymbol{\rho}_{z_{i}}^{\prime}\right)R\left(\boldsymbol{\rho}_{z_{i}}^{\prime\prime}\right)R^{\ast}\left(\boldsymbol{\rho}_{z_{i}}^{\prime\prime\prime}\right)\right\rangle \nonumber \\
 & =\delta\left(\boldsymbol{\rho}_{z_{i}}-\boldsymbol{\rho}_{z_{i}}^{\prime}\right)\delta\left(\boldsymbol{\rho}_{z_{i}}^{\prime\prime}-\boldsymbol{\rho}_{z_{i}}^{\prime\prime\prime}\right)+\delta\left(\boldsymbol{\rho}_{z_{i}}-\boldsymbol{\rho}_{z_{i}}^{\prime\prime\prime}\right)\delta\left(\boldsymbol{\rho}_{z_{i}}^{\prime}-\boldsymbol{\rho}_{z_{i}}^{\prime\prime}\right)
\end{align}
where $i=a,c$ and $\delta\left(\boldsymbol{\rho}-\boldsymbol{\rho}^{\prime}\right)$ is the delta function. The second term in Eq.~(\ref{eq: c_tau_nonvanishing}) is given by~\cite{shynk2012probability}
\begin{equation}\label{eq: autocorrelation_whiteNoise}
    \int_{0}^{\mathcal{T}}\text{d}t\left\langle N\left(t\right)N\left(t+\tau\right)\right\rangle =\delta\left(\tau\right).
\end{equation}
Upon substituting Eq.~(\ref{eq: field_profile}), (\ref{eq: field_at_PDs}), and (\ref{eq: cc_I}) -- (\ref{eq: autocorrelation_whiteNoise}) into Eq.~(\ref{eq: c_tau_nonvanishing}), we obtain after carrying out the integrations
\begin{equation}\label{eq: c_tau_result}
C\left(\boldsymbol{\rho}_{z_{b}},\tau\right)\propto \mathcal{T}\left|T\left(\boldsymbol{\rho}_{z_{b}}\right)\right|^{4}\cos\left(2\pi f_{mod}\tau\right)+\beta_{1}\beta_{2}\delta\left(\tau\right).
\end{equation}
It follows from Eq.~(\ref{eq: c_tau_result}) that the cross-spectrum is
\begin{align}\label{eq: cross_spectrum}
\Gamma\left(\boldsymbol{\rho}_{z_{b}},\omega\right) & =\int_{-\infty}^{\infty}C\left(\boldsymbol{\rho}_{z_{b}},\tau\right)e^{-i\omega\tau}\text{d}\tau\nonumber \\& =\Gamma_{0}\mathcal{T}\left|T\left(\boldsymbol{\rho}_{z_{b}}\right)\right|^{4}\delta\left(\omega-2\pi f_{mod}\right)+\beta_{1}\beta_{2},
\end{align}
where $\Gamma_{0}=\left(\frac{4\pi^2}{k^2}\right)^2\frac{(\pi\sigma^{2})^2\left|\tilde{A}\right|^{4}I^2_{0}}{f_{mod}}$. The cross-spectrum is a sum of frequency peak signal multiplied by $\left|T\left(\boldsymbol{\rho}_{z_{b}}\right)\right|^{2}$ and uniform noise background. Note the object (``1X'') can be represented by $\left|T\left(\boldsymbol{\rho}_{z_{b}}\right)\right|^{2}$. Scanning the object and recording $S_{1}\left(t\right)$ and $S_{2}\left(t\right)$ at every position $\boldsymbol{\rho}_{z_{b}}$, we can calculate the cross-spectrum as a function of the position. A heat map of $\Gamma\left(\boldsymbol{\rho}_{z_{b}},\omega=2\pi f_{mod}\right)$ will produce an image of the object,  because $\left|T\left(\boldsymbol{\rho}_{z_{b}}\right)\right|^{4}$ serves as a ``mask'' that modulates the amplitudes of the cross-spectrum from position to position as indicated in Eq. (\ref{eq: cross_spectrum}), and the shape of the object is finally encoded in $\Gamma\left(\boldsymbol{\rho}_{z_{b}},\omega=2\pi f_{mod}\right)$.
Furthermore, the larger the integration time $\mathcal{T}$ is, the greater the amplitude of the frequency peak will be; while, the noise is independent of $\mathcal{T}$. This suggests that, by increasing $\mathcal{T}$, the signal-to-noise ratio can be enhanced. Therefore, even though the output signal may undergo strongly scattering  and is below the noise level of the detectors, this method can still reconstruct the image of the target. 
\begin{figure}[ht]
	\centering
	\includegraphics[width=0.9\linewidth]{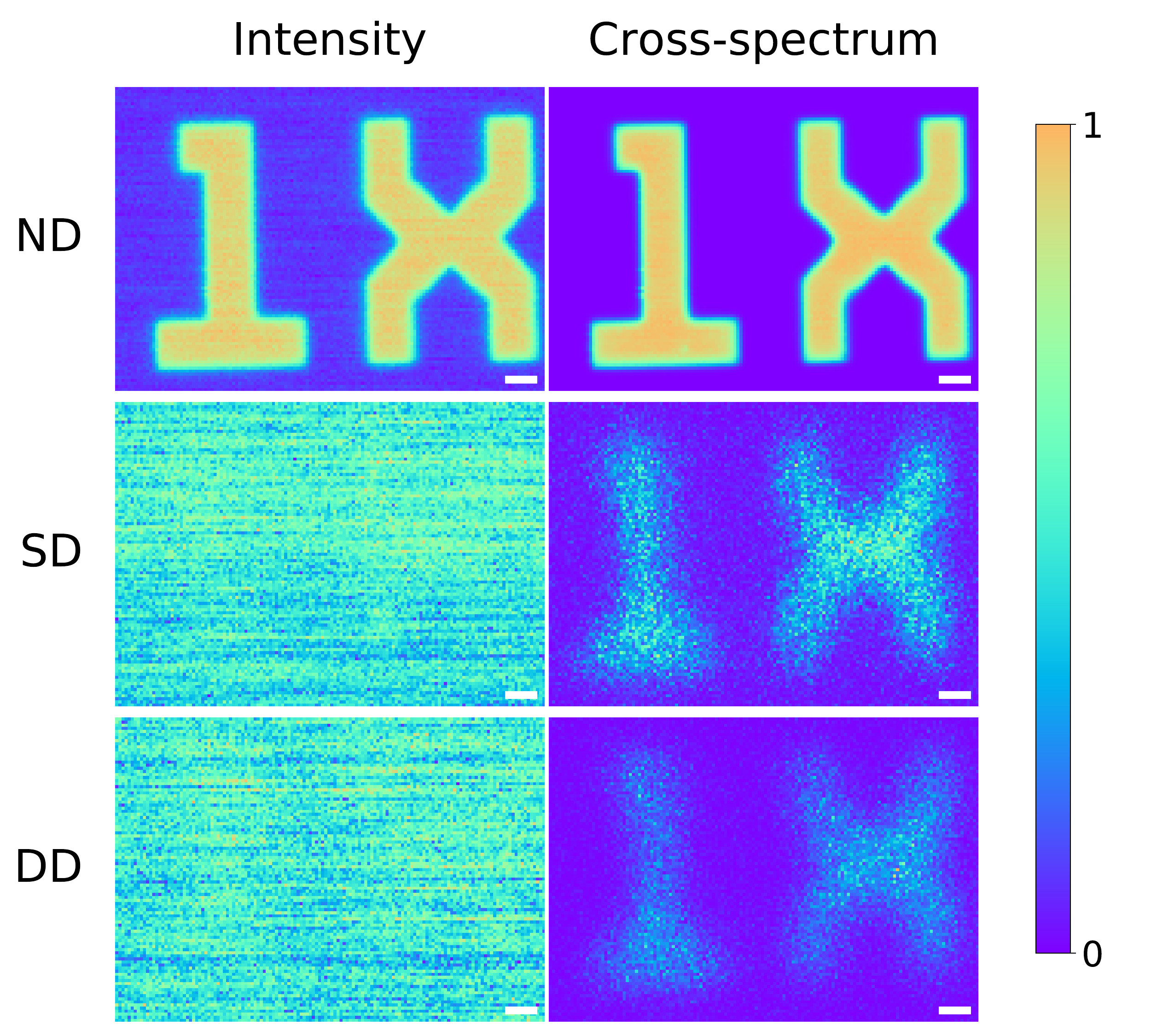}
	\captionsetup{justification=raggedright, singlelinecheck=false}
	\caption{Raster-scan images for an object with the letter "1X" being transparent and other regions being opaque.  Cross-spectrum images are generated by plotting $\Gamma\left(\boldsymbol{\rho}_{z_{b}},\omega=2\pi f_{mod}\right)$ (see Eq.~(\ref{eq: cross_spectrum})). Scale bar, 40 pixels. ND: no diffuser. SD: static diffuser. DD: dynamic diffuser.}
	\label{fig:Figure_2_Img}
\end{figure}

\section{Experimental results}\label{sec:experiments}
To demonstrate that our method works experimentally for both static and dynamic strongly scattering  media, we perform the measurements under three situations: 1. imaging without diffuser (no diffuser, ND); 2. the object is sandwiched between two static diffusers (SDs); and 3. the two diffusers are moved back and forth together by a motorized stage (dynamic diffusers, DDs). The speed and acceleration of the stage has a random number at every moment. The maximum value of the speed is 500 $mm/s$ and the maximum value of the acceleration is 500 $mm/s^2$. The full range is 1.5 cm. The object has the letter ``1X'' being transparent and other regions being opaque. For all three cases, data are collected by an oscilloscope with a fixed sample rate at 2 GHz. At each position, 1 million data points are taken to calculate the cross-spectrum, corresponding to 500 $\mu$s integration time which ensures a strong cross-correlation signal. The whole image contains $100\times 140$ pixels (number of steps scanned) with the pixel size (scanning step size) of 25 $\mu$m.
\begin{table}[htbp]
\caption{\label{tab:visibility_1} Visibility for different diffuser states}
    \begin{ruledtabular}
        \begin{tabular}{ccc}
         Diffuser State & Intensity & Cross-spectrum \\ \midrule
    		No diffuser (ND) & 0.725 & 0.967  \\ 
    		Static diffuser (SD) & 0.032 & 0.451 \\  
    		Dynamic diffuser (DD) & 0.031 & 0.558 \\
        \end{tabular}
    \end{ruledtabular}
\end{table}

\begin{table*}[htbp]
\caption{\label{tab:visibility_2} Visibility of different acquisition time for static and dynamic diffuser states}
    \begin{ruledtabular}
        \begin{tabular}{ccccc}

    		\multirow{2}*{Acquisition time} & \multicolumn{2}{c}{Static diffuser (SD)} & \multicolumn{2}{c}{Dynamic diffuser (DD)} \\ \cmidrule(lr){2-3} \cmidrule(lr){4-5}
    		& Intensity & Cross-spectrum & Intensity & Cross-spectrum \\
            \midrule
    		50 $\mu$s & 0.027 $\pm$ 0.004 & 0.135 $\pm$ 0.009 & 0.035 $\pm$ 0.005 & 0.190 $\pm$ 0.004 \\

    		100 $\mu$s & 0.033 $\pm$ 0.004 & 0.332 $\pm$ 0.001 & 0.030 $\pm$ 0.002 & 0.305 $\pm$ 0.004 \\

    		500 $\mu$s & 0.034 $\pm$ 0.003 & 0.466 $\pm$ 0.008 & 0.048 $\pm$ 0.004 & 0.560 $\pm$ 0.002 \\

    	\end{tabular}
    \end{ruledtabular}
\end{table*}

The main experimental result is shown in Fig.~\ref{fig:Figure_2_Img}, of which the pixel values $v$ are normalized by $\tilde{v}=(v-v_{min})/(v_{max}-v_{min})$. In the first column, we directly plot the intensity measured by the detectors; while, in the second column, we plot $\Gamma\left(\boldsymbol{\rho}_{z_{b}},2\pi f_{mod}\right)$. The first row shows ND images. The second and third row list images obtained with SD and DD, respectively.  We summarize the visibility of images in Table~\ref{tab:visibility_1}, which is calculated by $V=(\bar{\tilde{v}}_{s}-\bar{\tilde{v}}_{b})/(\bar{\tilde{v}}_{s}+\bar{\tilde{v}}_{b})$, where $\bar{\tilde{v}}_{s}$ and $\bar{\tilde{v}}_{b}$ are the average pixel values of signal (``1X'' region) and background, respectively. As shown in Fig.~\ref{fig:Figure_2_Img}, in both SD and DD cases, when the strongly scattering media is present, the recorded intensity does not show any image in either case, the extremely low visibility is a sign that our signal is truly at the noise level of the detectors. On the other hand, in both cases, the images are still retrieved using the cross-spectrum technique with high visibility. The results suggest that the cross-spectrum method can image an object hidden behind both static and dynamic strongly scattering media. We also notice that the visibility of the cross-spectrum image is higher than the intensity image even though there is no diffuser (ND), which suggests cross-spectrum is also an effective way to enhance signal-to-noise ratio when the scattering media is absent.
\begin{figure}[ht]
	\includegraphics[width=\linewidth]{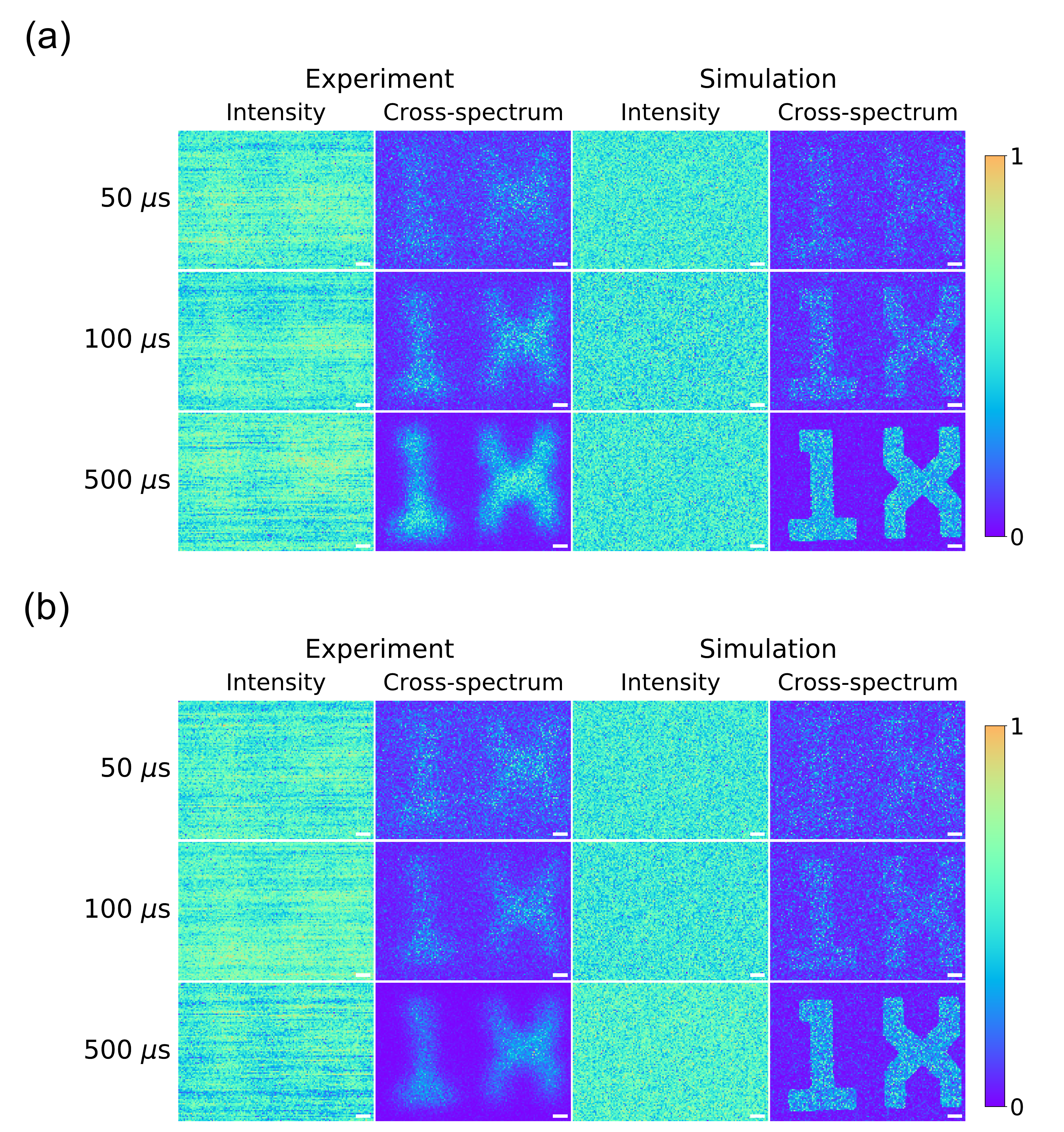}
	\captionsetup{justification=raggedright, singlelinecheck=false}
	\caption{Raster-scan images and simulations for different acquisition time with (a) static diffusers and (b) dynamic diffusers. Scale bar: 10 pixels (0.25~mm).}
	\label{fig: acq_len}
\end{figure}

To further test our method, we also compare the measured images from different acquisition time ($ 50~\mu s,100~\mu s$, and $500~\mu s$) with the same sample rate. The incident intensity of light is fixed for the static and dynamic diffusers. Simulation is also performed for the intensity and cross-spectrum based on Eq.~(\ref{eq: signal_at_detectors}) and Eq.~(\ref{eq: cross_spectrum}), respectively, as a comparison with the experimental results. The results are shown in Fig.~\ref{fig: acq_len}, the corresponding visibility of the experimental results are listed in Table \ref{tab:visibility_2}. It can be seen that, in general, the longer integration time is, the higher visibility one can achieve for both static and dynamic diffusers. This means we can obtain a clear image under strong scattering at the expense of long acquisition time. We note that the visibility of the recorded intensity image is kept extremely low even when one increases the acquisition time up to 10 times, nevertheless the cross-spectrum image becomes more and more clear. The visibility also increases much faster than that of the intensity measurement when increasing the acquisition time. We point out that the fundamental limit of imaging speed is the acquisition length, which is on the order of 100 $\mu$s for the current setup, but can be in principle orders faster with higher modulation frequency and higher sample rate (GHz range laser modulation speed and detection). The raster scan speed can also be much improved if, for instance, a 2D galvo-resonant scanner is integrated into the system. 

\section{Summary}
In conclusion, we have developed a cross-spectrum method to extract a weak optical signal from the extremely noisy background and image objects hidden behind scattering media. The major advantage of this scheme is that it uses CW laser with low power in a non-invasive manner which would be easy to implement and bio-tissue friendly. It suits for both static and dynamic media, which makes it adaptive in most application situations. Together with the fast acquisition time with current technology, our scheme paves the way for efficient imaging in previously inaccessible scenarios.

\section*{Acknowledgement} 
This research is supported by grants from: Air Force Office of Scientific Research (Award No. FA9550-20-1-0366 DEF), Office of Naval Research (Award No. N00014-20-1-2184), Robert A. Welch Foundation (Grant No. A-1261), National Science Foundation (Grant No. PHY-2013771), and Qatar National Research Fund (project NPRP 13S-0205-200258). The authors thank Y. J. Shen and T. Smith for helpful discussions.

\bibliography{imaging2}

\end{document}